# Facile synthesis and enhanced visible light photocatalytic activity of N and Zr co-doped $TiO_2$ nanostructures from nanotubular titanic acid precursors

Min Zhang*, Xinluan Yu, Dandan Lu and Jianjun Yang*

**Abstract**

Zr/N co-doped $TiO_2$ nanostructures were successfully synthesized using nanotubular titanic acid (NTA) as precursors by a facile wet chemical route and subsequent calcination. These Zr/N-doped $TiO_2$ nanostructures made by NTA precursors show significantly enhanced visible light absorption and much higher photocatalytic performance than the Zr/N-doped P25 $TiO_2$ nanoparticles. Impacts of Zr/N co-doping on the morphologies, optical properties, and photocatalytic activities of the NTA precursor-based $TiO_2$ were thoroughly investigated. The origin of the enhanced visible light photocatalytic activity is discussed in detail.

## Background

Recently, nanoscale $TiO_2$ materials have attracted extensive interest as promising materials for its applications in environmental pollution control and energy storage [1]. However, $TiO_2$ is only responsive to UV light ($\lambda <$ 380 nm, 3% to 5% solar energy) due to its large bandgap energy (typically 3.2 eV for anatase). It hinders the practical application of $TiO_2$ for efficient utilization of solar energy [2]. Many studies have been performed to extend the spectral response of $TiO_2$ to visible light and improve visible light photocatalytic activity by doping and co-doping with metals of V, Fe, Cu, and Mo or nonmetals of N, B, S, and C [3,4]. Among the efforts of mono-doping, nitrogen-doped $TiO_2$ was considered to be a promising visible light active photocatalyst. Asahi et al. reported that the effect of N doping into $TiO_2$ achieved enhanced photocatalytic activity in visible region than 400 nm [5]. Theoretical works revealed that the result of the narrowed bandgap is due to N doping-induced localized 2p states above the valence band [6]. However, these states also act as traps for photogenerated carriers and, thus, reduce the photogenerated current and limit the photocatalytic efficiency.

* Correspondence: zm1012@henu.edu.cn; yangjianjun@henu.edu.cn
Key Laboratory for Special Functional Materials of Ministry of Education, Henan University, Kaifeng 475004, People's Republic of China

In order to reduce the recombination rate of photogenerated carriers in the nitrogen-doped $TiO_2$, co-doping transition metal and N have been explored [7]. Recently, theoretical calculations have reported that visible light activity of $TiO_2$ can be even further enhanced by a suitable combination of Zr and N co-doping [8]. The Zr/N co-doping of anatase $TiO_2$ could narrow bandgap by about 0.28 eV and enhance the lifetimes of photoexcited carriers. Previously, we had fabricated N-doped $TiO_2$ with visible light absorption and photocatalytic activity using precursor of nanotubular titanic acid (NTA, $H_2Ti_2O_4$ $(OH)_2$) [9]. The visible light sensitization of N-doped NTA sample was due to the formation of single-electron-trapped oxygen vacancies (SETOV) and N doping-induced bandgap narrowing. It was also found that the N-doped $TiO_2$ prepared by NTA showed the highest visible light photocatalytic activity compared with the $TiO_2$ prepared by different other precursors such as P25 [10]. To obtain further enhanced photocatalytic performance, in this work, we prepared Zr and N co-doped $TiO_2$ nanostructures using nanotubular titanic acid (NTA) and P25 as precursors by a facile wet chemical route and subsequent calcination. A systemic investigation was employed to reveal the effects of Zr and N doping/codoping in the enhancement of visible light absorption and photoactivity of the co-doped $TiO_2$ made by NTA and P25. The results showed that Zr/N-doped $TiO_2$ nanostructures made by nanotubular NTA precursors show significantly enhanced visible light absorption and much higher photocatalytic performance than the Zr/N-doped P25 $TiO_2$ nanoparticles. This work provided a strategy for the further enhancement of visible light photoactivity for the $TiO_2$ photocatalysts in practical applications.

## Methods

### Synthesis of NTA precursors

The precursor of nanotubular titanic acid was prepared and used as a co-doped precursor according to the procedures described in our previous reports [11-13]. Briefly, the Degussa P25 $TiO_2$, a commercial standard $TiO_2$ photocatalyst, reacted with concentrated NaOH solution to obtain $Na_2Ti_2O_5 \cdot H_2O$ nanotubes, and then, NTA was synthesized by an ion exchange reaction of $Na_2Ti_2O_5 \cdot H_2O$ nanotubes with an aqueous solution of HCl.

### Preparation of N and Zr co-doped $TiO_2$

The as-prepared NTA was mixed with urea (mass ratio of 1:2) and dissolved in a 2% aqueous solution of hydrogen peroxide, followed by the addition of pre-calculated amount of $Zr(NO_3)_4 \cdot 5H_2O$ (Zr/Ti atomic ratio, 0%, 0.1%, 0.3%, 0.6%, 1.0%, 5.0%, and 10%). The resultant mixed solution was refluxed for 4 h at 40°C and followed by a vacuum distillation at 50°C to obtain the product of x% Zr/N-NTA. Final Zr/N co-doped $TiO_2$ were prepared by the calcination of x% Zr/N-NTA at a temperature range of 300°C to 600°C for 4 h. The target nanosized $TiO_2$ powder was obtained, denoted as x% Zr/N-$TiO_2$ (temperature), for example 0.6% Zr/N-$TiO_2$(500). For reference, Degussa P25 $TiO_2$ powders were used as precursor under the same conditions to prepare Zr/N co-doped $TiO_2$ (denoted as Zr/N-$TiO_2$(P25)).

### Characterization

The phase composition of various Zr/N co-doped $TiO_2$ samples were analyzed by X-ray diffraction (XRD, Philips X'Pert Pro X-ray diffractometer; Cu-Kα radiation, $\lambda = 0.15418$ nm). The morphologies of samples were observed using a transmission electron microscopy (TEM, JEOL JEM-2100, accelerating voltage 200 kV). Nitrogen adsorption-desorption isotherms were measured at 77 K on a Quantachrome SI automated surface area and pore size analyzer. The Brunauer-Emmett-Teller (BET) approach was used to evaluate specific surface area from nitrogen adsorption data. The UV-visible diffuse reflectance spectra (DRS) of the samples were obtained on a UV–vis spectrophotometer (Shimadzu U-3010, Kyoto, Japan) using $BaSO_4$ as the reference. The surface composition of the nanocatalysts was analyzed by X-ray photoelectron spectroscopy (XPS) on a Kratos Axis Ultra System with monochromatic Al Ka X-rays (1486.6 eV). An Axis Ultra X-ray photoelectron spectroscope (Quantera) was used for the chemical characterization of photocatalyst samples. The binding energies (BE) were normalized to the signal for adventitious carbon at 284.8 eV. The photoluminescence (PL) spectra were recorded on a fluorescence spectrometer (fluoroSE).

### Visible light photocatalytic activity

The photocatalytic activities of various Zr/N co-doped $TiO_2$ samples were evaluated by monitoring the oxidation process of propylene under visible light irradiation. About 25 mg of each photocatalyst sample was spread on one side of a roughened glass plate (*ca.* 8.4 $cm^2$ active area) and kept in a flat quartz tube reactor. A 300-W xenon lamp (PLS-SXE300/300UV, Beijing Trusttech Co. Ltd., China) was used as the visible light source. A cut filter ($\lambda \geq 420$ nm) was placed between the xenon lamp and reactor. The intensity of visible light irradiated on to be tested samples was *ca.*17.6 $mW \cdot cm^{-2}$. Pure $C_3H_6$ (99.99%) stored in a high-pressure cylinder was used as the feed gas, and the flow rate of the feed gas was adjusted to 150 mL/h. The concentration of $C_3H_6$, C was determined at a sensitivity of 1 ppm (*v/v* over volume) using a chromatograph (Shimadzu GC-9A) equipped with a flame ionization detector, a GDX-502 column, and a reactor loaded with Ni catalyst for the methanization of $CO_2$. The photocatalytic activity of visible light photocatalytic oxidation of $C_3H_6$ was calculated as $(C_0 - C)/C_0 \times 100\%$, where $C_0$ refers to the concentration of feed gas $C_3H_6$ feed gas.

## Results and discussion

Figure 1a shows the XRD patterns of Zr/N co-doped $TiO_2$ samples calcined at 500°C with various zirconium contents range from 0.1% to 10%. The diffraction peaks of all samples are ascribed to pure anatase phase (JCPDS: 21–1272), and no peaks assigned to oxides of zirconium were observed. The 2 theta values of 25.5°, 37.8°, 48.0°, 55.1°, and 62.7° correspond to anatase (101), (004), (200), (211), and (204) crystal planes, respectively [14]. The XRD results show that the Zr/N co-doped $TiO_2$ samples are anatase phase and confirm the absence of rutile and zirconia phase. It indicated that the zirconium species had been substituted into the crystal lattice sites of titania [15,16]. With increasing content of zirconium doping, the XRD peaks of all doped NTA samples exhibit significant peak broadening suggesting that the particle size of anatase $TiO_2$ decreased gradually. Figure 1b shows the XRD patterns of 0.6% Zr/N-$TiO_2$ samples calcined at 400°C, 500°C, and 600°C. The XRD intensity of anatase peaks becomes stronger and sharper with the increase of calcination temperature. There are no peaks assigned to oxides of zirconium, and rutile phase were observed even with 10% Zr content and the calcination temperature of 600°C. A similar phenomenon has been reported in Zr-doped $TiO_2$ system by Gao et al. [15]. They found that the Zr-doped $TiO_2$ sample containing even 20% Zr content exhibited only anatase phase and no signals of zirconium oxides presented when calcined at 500°C. They also claimed that the doping of Zr ions in $TiO_2$ lattice could reach about 30%. Recent reports show that the doping of zirconium in the lattice of $TiO_2$ prevented the anatase to rutile phase transformation during calcination [16-18]. Schiller et al. observed that Zr-doped $TiO_2$ showed a high phase stability and the anatase-type structure was maintained even after heat treatment at 800°C [18]. Here, we found similar results that rutile phase formation is suppressed with the co-doping of nitrogen and zirconium.

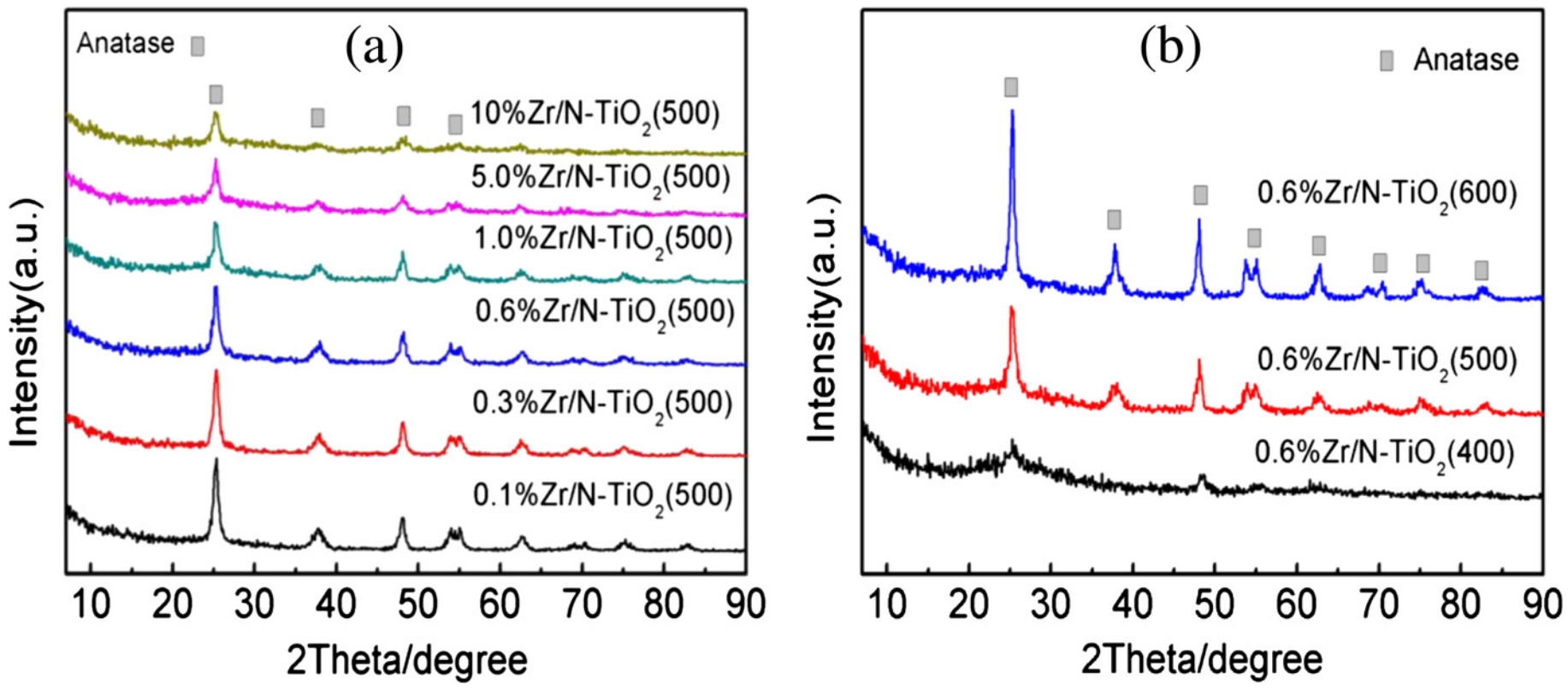

**Figure 1 XRD patterns of the samples. (a)** x%Zr/N-$TiO_2$(500), $x = 0.1, 0.3, 0.6, 1.0, 5.0, 10$; **(b)** samples of 0.6% Zr/N-$TiO_2$ calcined at 400°C, 500°C, and 600°C.

Figure 2 shows the typical TEM images of the prepared NTA precursor and 0.6%Zr/N-$TiO_2$ samples calcinated at 400°C, 500°C, and 600°C. Figure 2a shows the nanotubular morphology of NTA sample same with that reported in our previous results [11-13]. After the calcination in air at 400°C for 4 h, the 0.6%Zr/N-$TiO_2$ sample (Figure 2b) presented similar nanotubular morphology as that of the NTA precursor. In previous work, we found that the morphology of nanotubed $H_2Ti_2O_4(OH)_2$ could be easily destroyed when the calcination temperature was higher than 300°C [11]. The collapse of nanotube structure is due to the dehydration of inter-layered OH groups and crystallinity transition from orthorhombic system to anatase under calcination. In this work, the Zr/N co-doped NTA can still keep the nanotube structures with 400°C calcination. Figure 2c,d presents the 0.6% Zr/N-$TiO_2$ samples after thermal treatment at 500°C and 600°C. The nanotubular morphology of NTA precursor was changed to nanoparticles with high temperature calcination. Compared with the sample of 0.6% Zr/N-$TiO_2$(600) calcinated at 600°C, sample of 0.6% Zr/N-$TiO_2$(500) shows smaller pure anatase particles with size of *ca.* 10 nm and partially retained nanotubular structures. As we know, a smaller crystallite size, high surface area, and greater thermal stability are highly desirable properties for photocatalysts. Anatase type $TiO_2$ nanoparticles with small particle sizes (typically less than 10 nm) had exhibited enhanced photocatalytic activity because of the large specific surface area and quantum size effect [19,20]. In this work, better photocatalytic activity of 0.6% Zr/N-$TiO_2$ (500) sample was highly expected due to its pure anatase crystallinity and smaller crystallite size.

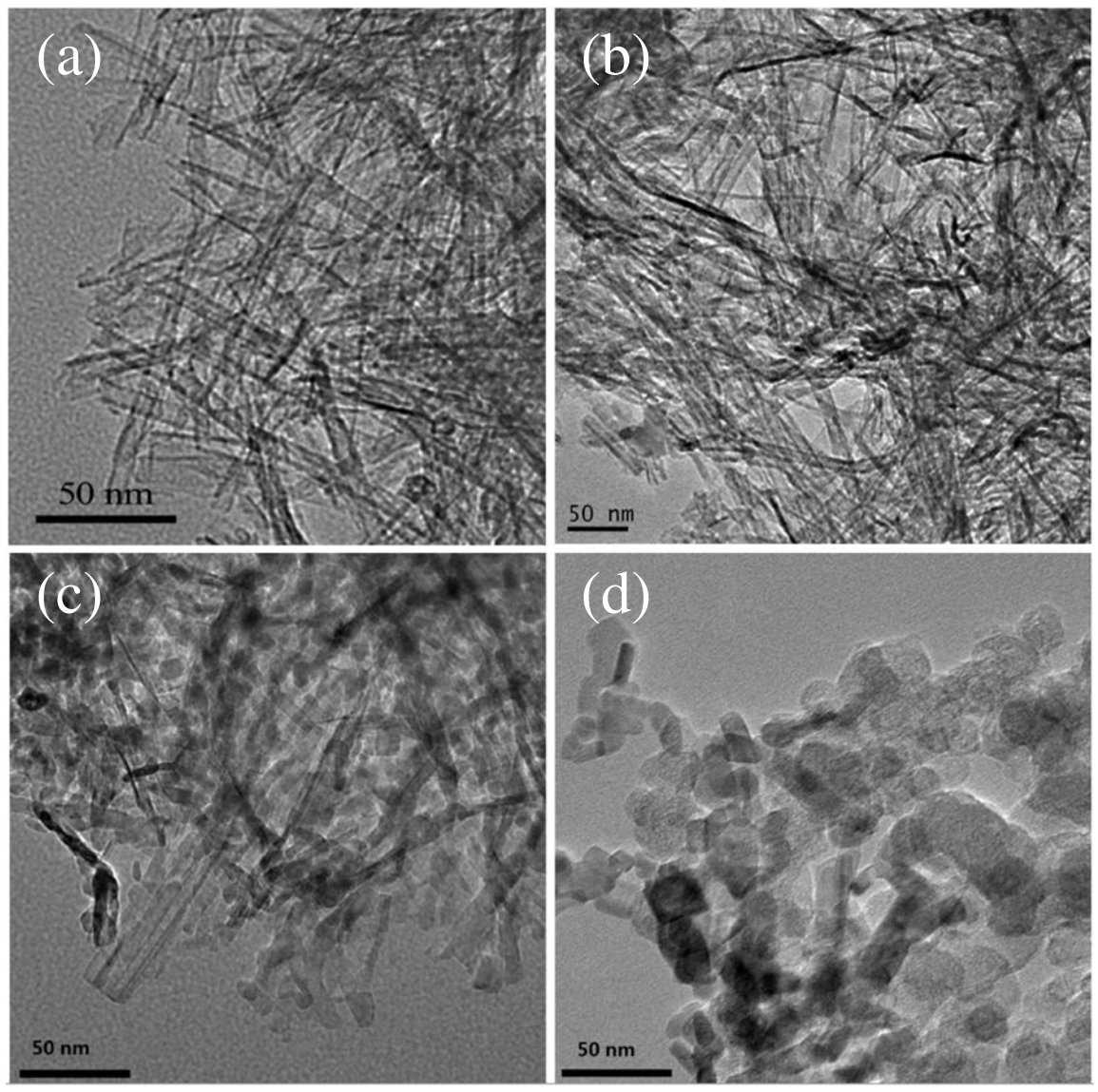

**Figure 2 TEM images of NTA precursor (a) and 0.6%Zr/N-$TiO_2$prepared at 400°C (b), 500°C (c), and 600°C (d).**

The surface areas of different doped samples measured by BET are shown in Tables 1 and 2. The BET results in Table 1 show that zirconium doping of x%-Zr-N-$TiO_2$-500 samples at the same calcination temperature exhibit an increase of specific surface area with increasing Zr content. This trend is due to the gradual decrease of crystallinity and particle sizes of anatase $TiO_2$ as demonstrated by XRD results in Figure 1a. The surface area data in Table 2 of 0.6%-Zr-N-$TiO_2$ samples calcined at different temperatures show a decreasing trend with the increase of calcination temperature. The XRD results in Figure 1b and TEM analysis in Figure 2 show that with increasing calcination temperature, the average crystallite size increases, in contrast with the BET surface areas that decrease.

**Table 1 BET surface areas of the x%-Zr-N-$TiO_2$-500 samples with different Zr doping concentration calcined at 500°C**

| Samples (x%-Zr-N-$TiO_2$-500) | Surface areas ($m^2g^{-1}$) |
|---|---|
| 0.1 | 122.31 |
| 0.3 | 142.96 |
| 0.6 | 143.04 |
| 1.0 | 166.25 |
| 5.0 | 218.18 |
| 10.0 | 240.18 |

Surface compositions of Zr/N co-doped $TiO_2$ samples were investigated by XPS. Figure 3a,b shows the high resolution XPS spectra of Ti 2p and O 1s for sample of 0.6% Zr/N-$TiO_2$(500). The binding energies of Ti $2p_{3/2}$ and Ti $2p_{1/2}$ components of 0.6% Zr/N-$TiO_2$(500) are located at 458.9 and 464.8 eV, corresponding to the existence of $Ti^{4+}$ state [11-13]. The O 1s spectrum in Figure 3b can be resolved into two peaks at 530.3 and 532.0 eV. The strong peak of 530.3 eV is ascribed to lattice oxygen in Ti-O bonds, and the small peak around 532.0 eV is ascribed to weakly physical adsorbed oxygen species such as $O^-$ and OH group on the surface [11-13]. The N 1s and Zr 3d spectra for samples of 0.6% Zr/N-$TiO_2$(500) can be observed in Figure 3c,d. The N 1s binding energy peaks are broad, extending from 396 to 403 eV. The center of the N1s peak locates at *ca.* 400.1 eV. In general, the assignment of the N 1s peak in the XPS spectra is under debate in the literature according to different preparation methods and conditions. We had attributed the N 1s peak at 400 eV to the interstitial N in the form of Ti-O-N in our previous reports [11-13]. Zr 3d peaks at 182.2 and 184.5 eV corresponding to the Zr $3d_{5/2}$ and Zr $3d_{3/2}$, respectively, are assigned to the $Zr^{4+}$ state of zirconium [16]. The above XPS results indicate that both nitrogen and zirconium are doped into the $TiO_2$ samples after calcination at 500°C.

Optical absorption properties of precursors (P25 and NTA), Zr doped and Zr/N co-doped P25 and NTA were studied by the diffuse reflectance in visible light region. Figure 4 shows the UV–vis DRS of prepared samples in the range of 400 to 700 nm. The undoped sample of P25 and NTA shows no visible light absorption. Zirconium mono-doped NTA sample also presents no obviously visible light absorption. It indicates that zirconium mono-doping may not lead to the bandgap narrowing of $TiO_2$ with NTA as precursor. Theoretical studies had proved that Zr mono-doping did not change the bandgap of $TiO_2$ and eventually did not exhibit better absorption ability in visible light region [8]. However, the spectra of Zr/N co-doped NTA shows a significantly broader absorption shifted to the visible region. While the absorption edge of Zr/N co-doped P25 sample only gets a slight shift to the visible region. The significant visible light absorption of Zr/N NTA indicates that the NTA is a better candidate than P25 as a precursor for N doping. We had reported the effect of annealing temperature on the morphology, structure, and photocatalytic behavior of NTA precursor [11]. The NTA experienced the process of dehydration and crystallinity transition during calcination, which is clearly beneficial for the N doping into the lattice of $TiO_2$. Moreover, single-electron-trapped oxygen vacancies (SETOV) were generated in the dehydration process [11]. In a recent study of visible light absorption and photocatalytic activity of N doped NTA, we demonstrated that the absorption shift to the visible light region of N-NTA samples is ascribed to the formation of single-electron-trapped oxygen vacancies (SETOV) in $TiO_2$ matrix and nitrogen doping [15]. In present work, zirconium mono-doping was found not to effectively narrow the bandgap of $TiO_2$. Herein, we also attributed the visible light absorption of Zr/N co-doped NTA to formation of SETOV and N doping.

The separation efficiency of photogenerated electron and hole is an important factor to influence the photocatalytic activity of $TiO_2$ samples. A lower recombination rate of photogenerated electron and hole is expected for higher photocatalytic activity. In order to examine the recombination rate of charge carriers, PL measurements were performed for the Zr/N-doped $TiO_2$ nanostructures made by NTA precursors. Figure 5 shows the PL emission spectra of undoped $TiO_2$ and Zr/N-doped $TiO_2$ with different zirconium contents under a 380-nm excitation. Obvious emission peaks at *ca.* 495 and 600 nm and a weak shoulder peak at 470 nm are observed for all samples. The peaks around 470 and 495 nm corresponds to the charge transfer transition from oxygen vacancies trapped electrons [21], while the peaks of 600 nm are attributed to the recombination of self-trapped excition or other surface defects [22]. As shown in Figure 5, the PL intensity of Zr/N-$TiO_2$ samples with Zr doping is lower than that of the pure NTA sample. It indicates that the Zr/N doping can efficiently inhibit the charge transfer transition from oxygen vacancies trapped electrons. The PL intensity of Zr/N-$TiO_2$ samples with lower Zr doping concentration shows a decreasing trend in the range of 0.1% to 1%. The low emission intensity associated with expected high photocatalytic activity is observed in the spectrum of 0.6% to 1% Zr/N-$TiO_2$ (500) samples. With more Zr doping such as 5%, the PL intensity of Zr/N-$TiO_2$ sample started to increase again. Finally, the 10%-Zr/N-$TiO_2$ sample has the highest intensity compared to other doped samples, which shows the excess doping of Zr ions into $TiO_2$ lattice introduced more recombination centers.

**Table 2 BET surface areas of the 0.6%-Zr-N-$TiO_2$samples calcined at different temperatures**

| Calcination temperature (°C) | Surface area ($m^2g^{-1}$) |
|---|---|
| 400 | 320.54 |
| 500 | 143.04 |
| 600 | 112.01 |

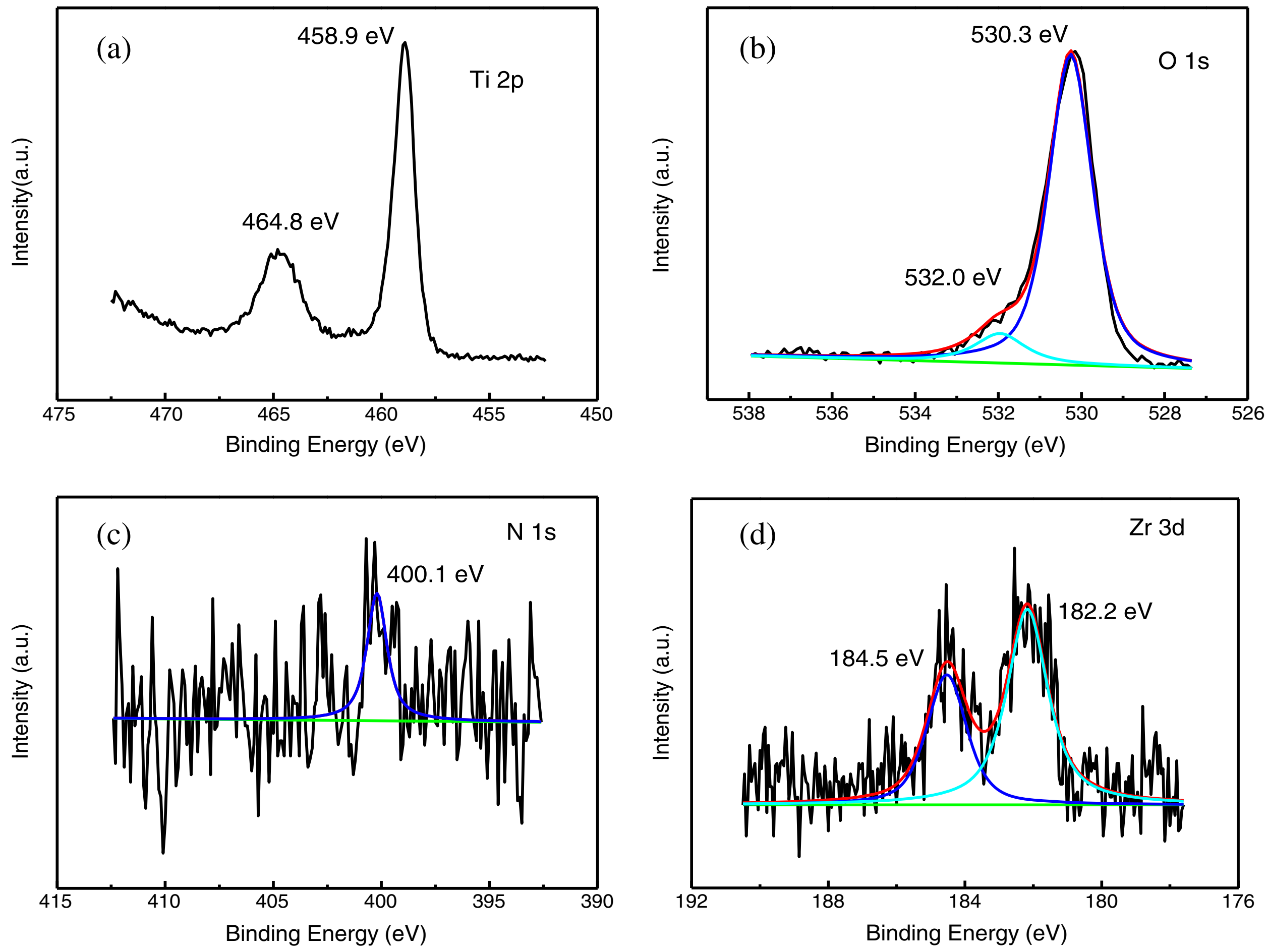

**Figure 3 High-resolution XPS spectra of Ti 2p (a), O 1 s (b), N 1 s (c), and Zr 3d (d) for sample of 0.6% Zr/N-$TiO_2$(500).**

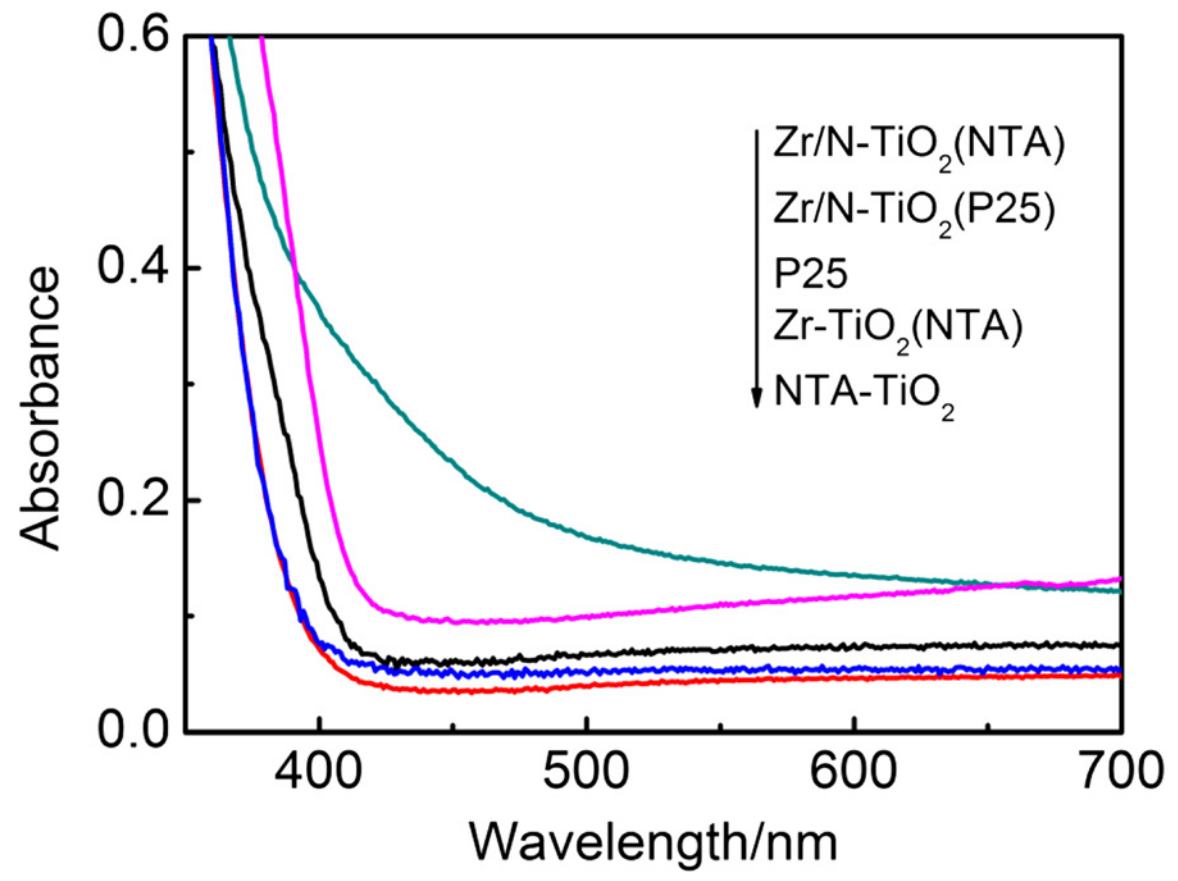

**Figure 4 UV–vis absorption spectra of precursor (P25 and NTA), Zr-doped NTA and Zr/N co-doped samples (P25 and NTA).** Prepared at 500°C with 0.6% Zr content.

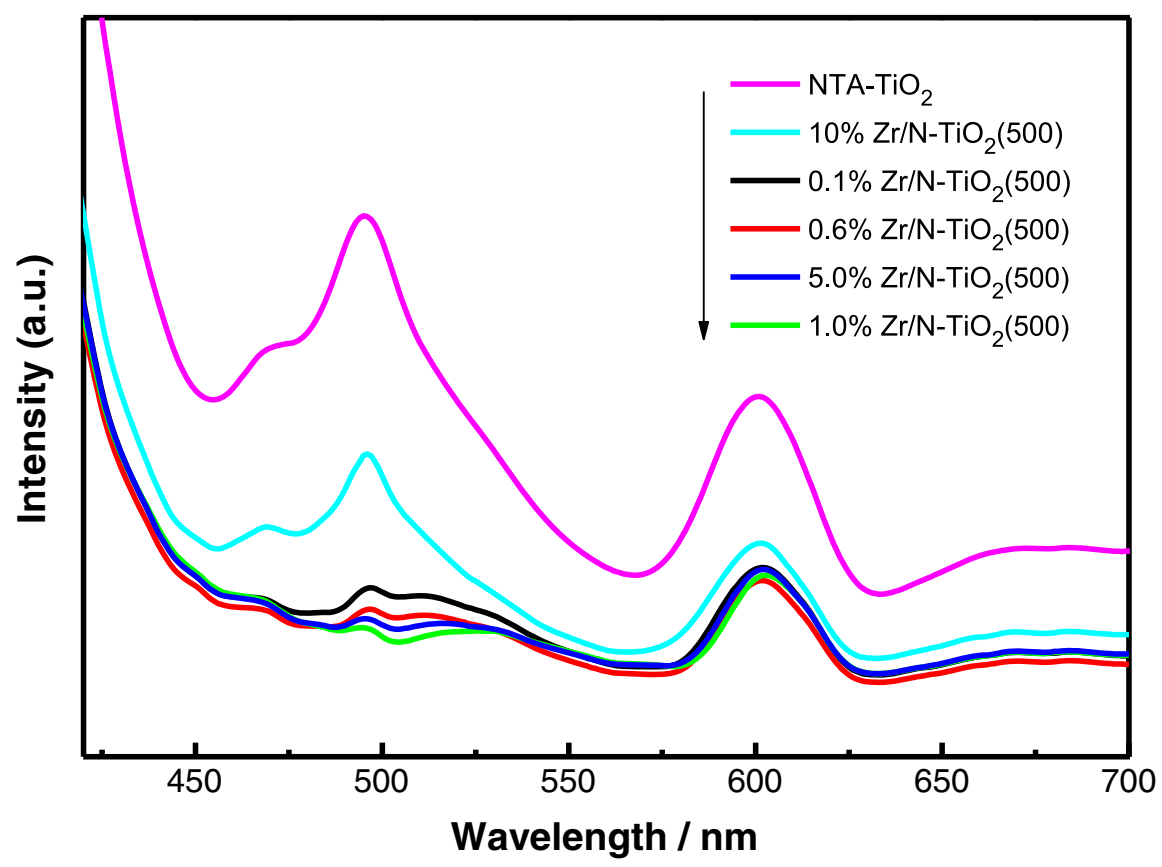

**Figure 5 PL spectra of as prepared samples with different Zr content ($\lambda_{ex} = 380$ nm).**

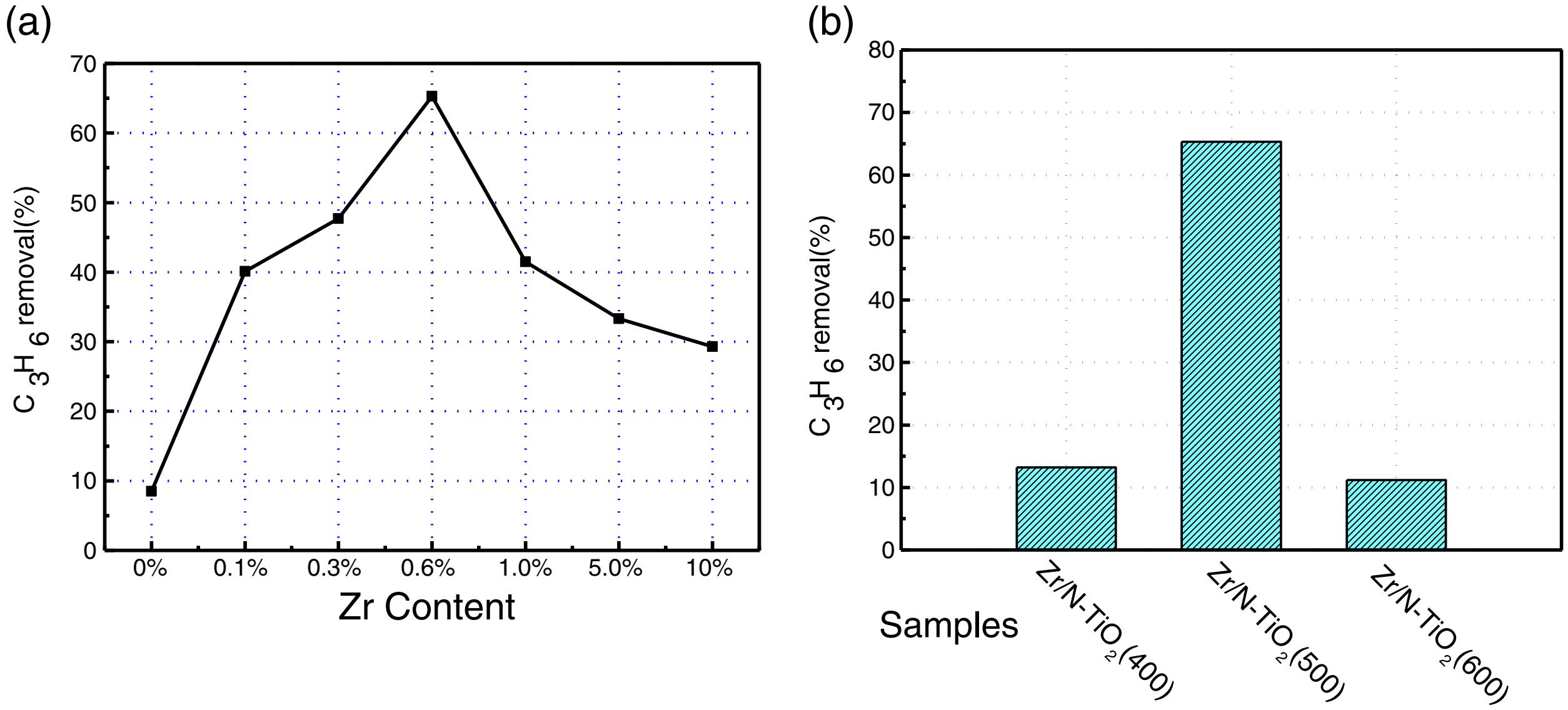

**Figure 6 Degradation of propylene over as-prepared catalysts under visible light irradiation. (a)** x% Zr/N-$TiO_2$(500), $x = 0$ to 10; **(b)** 0.6% Zr/N-$TiO_2$ calcined at 400°C, 500°C, and 600°C.

The photocatalytic activities of a series of prepared Zr/N co-doped NTA samples were investigated by photocatalytic oxidation of propylene under visible light irradiation. Figure 6a shows the visible light photocatalytic performance of $C_3H_6$ removal for Zr/N co-doped NTA samples with various zirconium doping amounts after 500°C calcination. The single N doped sample of N-$TiO_2$ (500) with 0% zirconium content shows a low visible light photocatalytic activity of *ca.* 10%. With the increase of zirconium content, the Zr/N-$TiO_2$ (500) samples show sharply increased photocatalytic activities. The best removal rate of propylene is found to be 65.3% for the 0.6%Zr/N-$TiO_2$ (500) sample. Then, the removal rate is decreased to about 30% with the increased zirconium doping amount up to 10%. It indicates that there is optimal amount for zirconium doping to get higher photocatalytic activity under visible light irradiation.

Figure 6b shows the visible light photocatalytic activities of 0.6% Zr/N-$TiO_2$ samples calcined at different temperatures. The 0.6%Zr/N-$TiO_2$ (400) sample calcined at 400°C shows a lower removal rate of *ca.* 12%. This lower photocatalytic activity is due to its poor anatase crystallinity as shown in XRD results. Compared with the 0.6% Zr/N-$TiO_2$ (600) sample, 0.6% Zr/N-$TiO_2$(500) sample shows the highest removal rate of *ca.* 65%. We considered the best photocatalytic performance of Zr/N-$TiO_2$(500) that is due to its higher crystallinity and high surface area according to the above XRD and TEM analysis.

For comparison, Degussa P25 was also used as a precursor to prepare doped $TiO_2$ samples. The photocatalytic activity of all $TiO_2$ samples were investigated under visible light irradiation after N mono-doping and Zr/N co-doping. Figure 7 shows the removal rate of N mono-doped and Zr/N co-doped samples made from precursors of P25 and NTA after 500°C calcination. For N mono-doping, the removal rate of N-doped P25 is 3% and the value increased to 12% for N-doped NTA-$TiO_2$. We had compared the visible light photocatalytic activities of N-doped $TiO_2$ made by different precursors such as P25 and NTA [9]. The highest photocatalytic performance was found for N-doped $TiO_2$ using NTA as precursor. In the Zr/N co-doping system, the removal rate of Zr/N-P25 is 9%, whereas the value of 0.6%Zr/N-NTA (500) increased to 65.3%.

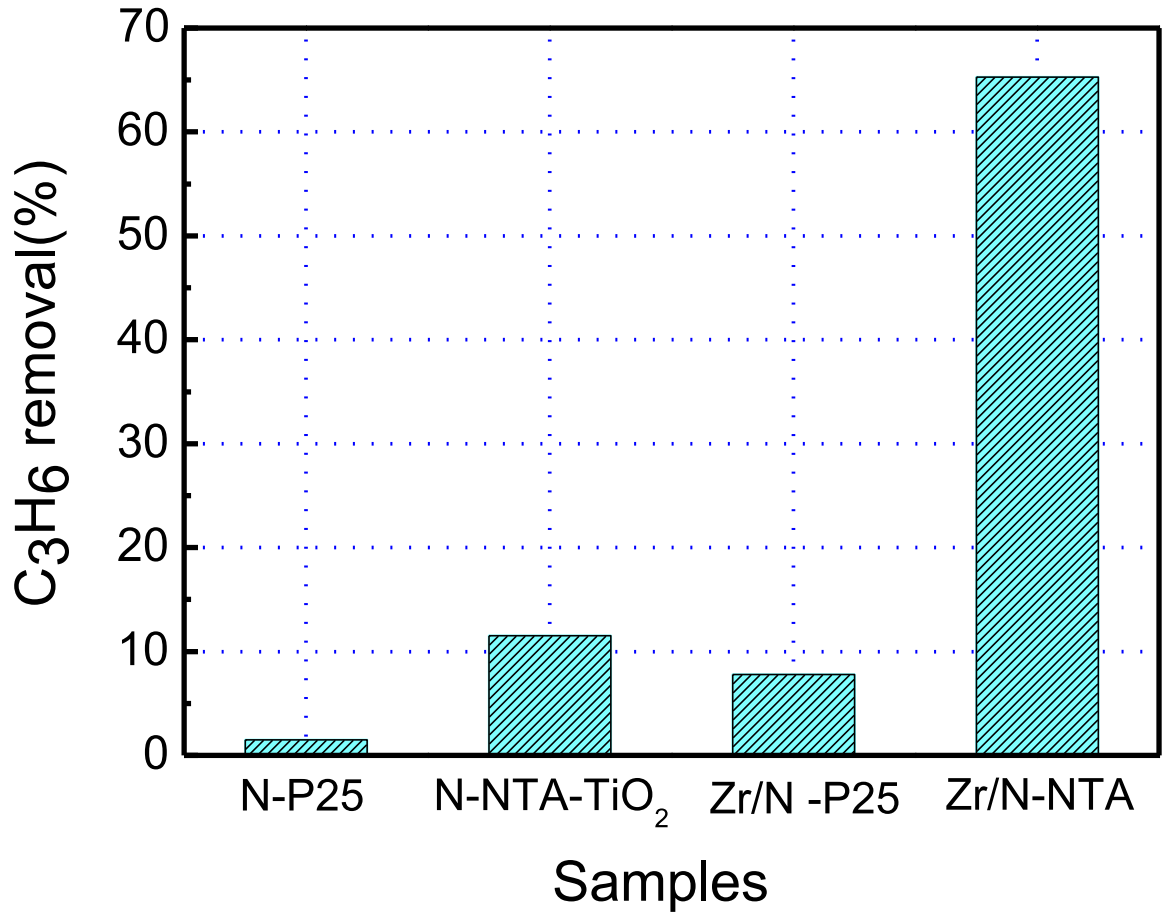

**Figure 7 Degradation of propylene over 0.6% Zr/N-$TiO_2$(500) synthesized from NTA and P25 respectively, as well as the N-NTA-$TiO_2$and N-P25.**

The results showed that the Zr/N codoping significantly enhanced the visible light photocatalytic activities of $TiO_2$ made by NTA precursor. It proves that NTA is a good candidate as a precursor for the preparation of promising visible light $TiO_2$ photocatalyst. As a special structural precursor, the process of loss of water and crystal structural transition during the calcination of NTA is expected to be beneficial for Zr and N doping into the lattice of $TiO_2$. Previously, the visible light absorption and photocatalytic activity of N-doped $TiO_2$ sample N-NTA was found to co-determine by the formation of SETOV and N doping induced bandgap narrowing [9]. Zr doping did not change the bandgap of $TiO_2$ and exhibit no effect on the visible light absorption in our experiments. However, theoretical calculation showed Zr doping brought the N 2p gap states closer to valence band, enhancing the lifetimes of photogenerated carriers [8]. Moreover, Zr doping effectively suppressed the crystallite growth of nano-$TiO_2$ and anatase to rutile phase transformation according to XRD and TEM analysis. Compared with Zr/N-P25, Zr/N-NTA(500) has the advantage of smaller crystallite size, larger surface area, and higher concentration of Zr and N dopant. It is shown that enhanced photocatalytic activity of Zr/N-NTA is achieved in the visible light region as a result of synergy effect of N/Zr codoping and use of nanotubular NTA as precursor.

## Conclusions

In summary, Zr/N co-doped $TiO_2$ nanostructures were successfully synthesized using nanotubular titanic acid (NTA) as precursors by a facile wet chemical route. The Zr/N-doped $TiO_2$ nanostructures made by NTA precursors show significantly enhanced visible light photocatalytic activities for propylene degradation compared with that of the Zr/N co-doped commercial P25 powders. Impacts of Zr/N co-doping on the morphologies, optical properties, and photocatalytic activities of the NTA-based $TiO_2$ were thoroughly investigated to find the origin of the enhanced visible light active photocatalytic performance. It is proposed that the visible light response is attributed to the intra-band by the nitrogen doping and calcination-induced single electron-trapped oxygen vacancies (SETOV). Crystallization and growth of Zr/N-doped $TiO_2$ were also impacted by the addition of zirconium. The best visible light photocatalytic activity of Zr/N co-doped NTA was achieved by co-doping with optimal dopant amount and calcination temperature. This work also provided a new strategy for the design of visible light active $TiO_2$ photocatalysts in more practical applications.

#### Competing interests

The authors declare that they have no competing interests.

#### Authors' contributions

XLY carried out the synthesis, characterization, and photocatalytic degradation experiments. DDL participated in the synthesis and TEM characterization experiments. MZ and JJY participated in the design and preparation of the manuscript. All authors read and approved the final manuscript.

#### Acknowledgements

The authors thank the National Natural Science Foundation of China (no.21203054) and Program for Changjiang Scholars and Innovation Research Team in University (no. PCS IRT1126).

Received: 18 November 2013 Accepted: 19 December 2013
Published: 26 December 2013